\documentclass[prd,preprint,tightenlines,showpacs,preprintnumbers,nofootinbib,eqsecnum,superscriptaddress]{revtex4}

 \usepackage[dvips,final]{graphicx}
  \usepackage{amssymb}
   \usepackage{amsmath}
    \usepackage{amsfonts}
     \usepackage{epsfig}
      \usepackage{bm}

\usepackage[section]{placeins}

\usepackage{ctable}
\usepackage{booktabs}
\usepackage{array}
\usepackage{tabularx}
\usepackage{xcolor}
\usepackage{pstricks}

\begin{document}


\title{A new parton fragmentation procedure \\ 
for heavy hadron production in proton-proton collisions}

\author{Antoni Szczurek}
\email{antoni.szczurek@ifj.edu.pl}
\affiliation{Institute of Nuclear Physics PAN, ul. Radzikowskiego 152,
PL-31-342 Krak\'ow, Poland \\
College of Natural Sciences, Institute of Physics,
University of Rzesz\'ow, ul. Pigonia 1, PL-35-310 Rzesz\'ow, Poland\vspace{1cm}}

\begin{abstract}
We propose a new procedure for parton-to-hadron fragmentation in
proton-proton collisions. The hadronization is considered in the
parton-parton center-of-mass system and hadrons are assumed to be
emitted in the direction of outgoing partons in that frame and then 
their four-momenta are transformed to the overall center-of-mass system 
using relevant Lorentz transformations and appropriate distributions 
are constructed.
For heavy hadron production the energy-momentum condition is imposed.
In many cases our procedure disagrees with the commonly used rule 
that rapidity of produced hadron is the same as rapidity of the parent parton.
We illustrate the newly proposed scheme for production of $D$ mesons,
$\Lambda_c$ baryon and $\eta_c$ quarkonium in leading-order approach
to parton production. 
The transverse momentum, rapidity and Feynman-$x_F$ distributions are shown.
We consider $c \to D$ and $g \to D$ as well as
$c \to \eta_c$ and $g \to \eta_c$ hadronization processes.
The resulting rapidity distributions are much narrower than those
obtained in the traditional approach with parton-hadron rapidity
equivalence. This has consequences both for mid- and forward rapidities
and could have consequences for high-energy neutrino production in 
the Earth's atmosphere.
We discuss also $c \to \Lambda_c^+$ and $b \to \Lambda_c^+$
fragmentation and find that the second mechanism could be partly
responsible for the enhanced production of $\Lambda_c^{\pm}$ observed
recently by the ALICE collaboration.

\end{abstract} 

\pacs{12.38.Bx, 13.85.Ni, 14.40.Pq}
\maketitle

\section{Introduction}

The hadronization process and production of a given hadron is often done
in terms of the parton-to-hadron fragmentation functions.
In $e^+ e^-$ collisions the whole $e^+ e^-$ CM energy goes to
production of particles. In leading-order the quark and antiquark are
emitted back-to-back. The results are presented as
$\frac{d \sigma}{d x_E}$, where $x_E$ is momentum or energy
fraction of the initial electron/positron.
The theoretical aspects of heavy flavour fragmentation functions 
in $e^+ e^-$ collisions were discussed e.g. in \cite{NO2000,CNO2005}.

In proton-proton collisions rather rapidity and/or transverse momentum
distributions of hadrons of interest are presented. It is usually
assumed that rapidity of the hadron is the same as the rapidity of
parent parton. This is a standard procedure for $c \to D$ or $b \to B$
hadronization (see e.g. \cite{Cacciari2002}), although small modifications 
were considered e.g. in \cite{MS2020_fragmentation}.
The standard approach is used e.g. in often used code FONLL
\cite{FONLL}.

In proton-proton collisions only a small part of initial energy is used
in the parton fragmentation process.
A big part of energy is related to so-called remnant fragmentation 
where the underlying physics is somewhat different.

We are particularly interested in production of heavy hadrons, both
mesons and baryons. The $D$ and $B$ meson production are classical
examples.
As discussed e.g. in \cite{MS2020_fragmentation} the $y_Q = y_H$
assumption used commonly in this contex has limitations, especially
for forward production of heavy hadrons. The forward production
of $D$ mesons is difficult experimentally epecially at the LHC because
there is no instrumentation in this part of the phase space.

The forward production of $D$ mesons is, however, very important in the
context of high-energy neutrino production in the Earth's atmosphere.
The IceCube collaboration measured neutrinos with energies $E_{\nu}
\sim$ 10$^7$ GeV \cite{IceCube}. It was claimed that such neutrinos 
are extra terrestial and even extra galactic.
However, a big part of such neutrinos could be produced in the
collisions of high-energy cosmic rays in the atmosphere 
(see e.g. \cite{GMPS2017,BGKKMS2017} and references therein).
How big is this part is not fully understood in our opinion and may
strongly depend on the model of $c \to D$ fragmentation.
Forward production of neutrinos at accelerators is another problem
of interest \cite{MSZB2020} (fixed target mode), \cite{BDGJH2020}
(collider mode) in this context.

In the present paper we consider a simple model which takes
into account the partonic subprocess as a basis for the underlying dynamics.

The production of heavy exotic hadrons is a new and interesting topic
of growing interest at the LHC.
For example the LHCb collaboration measured production of pseudoscalar
$\eta_c(c \bar c)$ meson \cite{Aaij:2014bga}. The dominant production 
mechanism seems to be the color-singlet gluon-gluon fusion 
\cite{Baranov:2019joi,BPSS2020_etac} but fragmentation mechanisms are also 
possible (see e.g.\cite{Nejad2015}).
We shall discuss this case in more detail in the present paper.

Recently the LHCb collaboration discovered one of the $\Xi_{ccq}$
baryons containing two charm quarks \cite{LHCb_Xicc}. 
It may be expected that soon their
results will be presented as normalized cross section and normalized
distributions in rapidity and transverse momentum.
Different mechanisms for its production were considered in the literature
\cite{MS1996,Baranov1996,Baranov1997,GS2002,CQWW2006}, including fragmentation 
mechanism(s). 

The possible observation of triply charmed $\Omega_{ccc}$ baryon is
currently under discussion \cite{S1999,BS2004,CW2011,WX2018}.

In the present paper we wish to revise the simple model of
$Q \to H_{Q}$ and $g \to H_{Q}$ fragmentation used so far in a
too-simplified way, at least in our opinion.

We consider a general situation of parton to heavy hadron fragmentation.

We start from heavy quark/antiquark fragmentation
\begin{equation}
Q \to H_{Q}  \; \; \; or \; \; \;  {\bar Q} \to \overline{H}_Q   \; , 
\label{heavy_quark_fragmentation}
\end{equation}
where $Q/{\bar Q}$ are heavy quark/antiquark and $H_Q / {\overline H}_{Q}$
are hadrons (mesons or baryons) that contain $Q$ or $\bar Q$.\\
Interesting examples are:
\begin{eqnarray}
c &\to& D^0, D^+, D_s^{+} \; , \nonumber \\
c &\to& \Lambda_c^+, \Xi_{ccq}, \Omega_{ccc}  \; , \nonumber \\
c &\to& \eta_c, \chi_c, J/\psi  \; .
\end{eqnarray}
The scheme considered in the present paper can be also used for
\begin{equation}
q \to {\bar D}(\bar c q) \;\;\; or \;\;\; {\bar q} \to D(c \bar q)
\label{subleading_fragmentation}
\end{equation}
subleading fragmentation considered e.g. in the context of $D^{\pm}$ meson
production asymmetry \cite{MS2018_asymmetry} or for
\begin{equation}
g \to D/{\bar D}, \Lambda_c^{\pm}, \eta_c, J/\psi, 
\Omega_{ccc}/{\bar \Omega}_{ccc}
\end{equation}
considered in the literature \cite{KK2005,Nejad2015,KKSS2020}.

\section{Parton to heavy hadron fragmentation
in proton-proton collisions}

Here we discuss how to calculate the fragmentation process
in the case of production of very heavy object from
much lighter parton (gluon or charm quark/antiquark) fragmentation. 
 
In the standard fragmentation function approach the meson/baryon
transverse momentum ditribution is calculated as:
\begin{equation}
d\sigma/dp_{t,H}(p_{t,H}) = \int dz \; D(z) \; d\sigma/dp_{t,Q}(p_{t,Q}) \; ,
\end{equation}
where $d\sigma/dp_{t,Q}$ is transverse momentum distribution of 
heavy quark/antiquark and
\begin{equation}
p_{t,H} = z p_{t,Q} \; .
\end{equation}
The cross section is calculated taking into account gluon-gluon fusion  
$g g \to Q \bar Q$ or quark-antiquark annihilation $q \bar q \to Q \bar Q$.
This approach does not include energy-mementum conservation
in the parton subsystem
which is important in some regions of the phase space.
We shall discuss how to improve the standard fragmentation approach
to proton-proton scattering.
For illustration we shall concentrate on leading-order collinear
approach. We hope that in future the discussed approach can be extended
to higher-orders and $k_t$-factorization approach.

As an example we consider also production of relatively heavy 
$\eta_c(c \bar c)$ production in hadronization.
The $p p \to \eta_c c \bar c X$ process is considered recently in
non-relativistic QCD approach \cite{Baranov2020}.
For the reaction $p p \to \eta_c c \bar c X$ there are
three emitted particles and
\begin{eqnarray}
x_1 &=& \frac{\sqrt{p_{t,\eta_c}^2 + m_{\eta_c}^2}}{\sqrt{s}} \exp(+y_{\eta_c})
      + \frac{\sqrt{p_{1t}^2 + m_c^2}}{\sqrt{s}} \exp(+y_1)
      + \frac{\sqrt{p_{2t}^2 + m_c^2}}{\sqrt{s}} \exp(+y_2)   \; , \nonumber \\
x_2 &=& \frac{\sqrt{p_{t,\eta_c}^2 + m_{\eta_c}^2}}{\sqrt{s}} \exp(+y_{-\eta_c})
      + \frac{\sqrt{p_{1t}^2 + m_c^2}}{\sqrt{s}} \exp(-y_1)
      + \frac{\sqrt{p_{2t}^2 + m_c^2}}{\sqrt{s}} \exp(-y_2)   \; .
\label{x_values_2to3}
\end{eqnarray}
In the fragmentation picture one $c$ or $\bar c$ is not explicit
and therefore usually not included in the evalution of $x_1$ and $x_2$,
i.e.
\begin{eqnarray}
x_1 &=& \frac{p_t}{\sqrt{s}} \exp(+y_1) + \frac{p_t}{\sqrt{s}}
\exp(+y_2) \; , \nonumber \\
x_2 &=& \frac{p_t}{\sqrt{s}} \exp(-y_1) + \frac{p_t}{\sqrt{s}}
\exp(-y_2) \; .
\label{x_values_2to2} 
\end{eqnarray}
Clearly $x_1^{real} > x_1^{stand}$ and $x_2^{real} > x_2^{stand}$.
So the standard FF approach overestimates the cross section for $\eta_c$
production in $c/{\bar c} \to \eta_c$ fragmentation.

In the following we discussed how to improve the standard LO
collinear-factorization calculation for heavy hadron production
to be used in the full phase space.

\begin{figure}
\includegraphics[width=7cm]{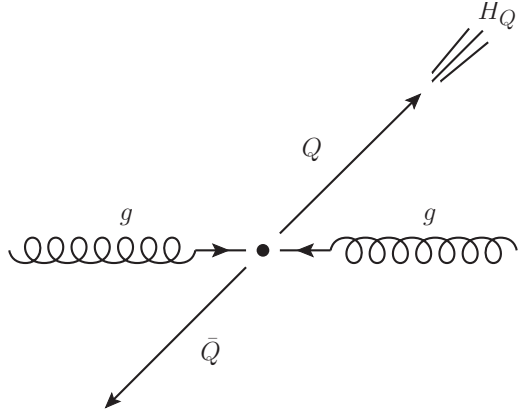}
\caption{The view of the fragmentation in the $Q \bar Q$ frame
of reference.}
\label{fig:diagram}
\end{figure}

Let us describe proposed by us fragmentation procedure.

\begin{itemize}

\item First the $\frac{d \sigma}{d y_1 d y_2 d p_t}$ three-dimensional
distributions of parton rapidities and transverse momenta are
calculated, i.e. we have full kinematical information about
$Q (p_1=(E_1,\vec{p}_1))$ and ${\bar Q} (p_2=(E_2,\vec{p}_2))$.

\item Then $Q {\bar Q}$ rest frame is found i.e. we can calculate
velocity of the $Q \bar Q$ system $\vec{\beta}_{pair}$
in the overall center-of-mass system.

\item The momenta of $Q$ and ${\bar Q}$ are next transformed to the rest
  frame of the diparton system (see Fig.\ref{fig:diagram})
\begin{eqnarray}
p_1 &\to& p_{1/pair}   \; , \nonumber \\
p_2 &\to& p_{2/pair}   \; .
\label{CM_to_RCM_transformation}
\end{eqnarray}
The relevant formulae can be found e.g. in Ref.\cite{Hagedorn}.
Then, in this specific system we have:
\begin{equation}
E_{1/pair} = E_{2/pair} = \frac{\sqrt{{\hat s}}}{2} \;\;\; and  \;\;\; 
\vec{p}_{1/pair} = - \vec{p}_{2/pair} \; .
\label{in_RCM}
\end{equation}
This is in full analogy to what is done in the $e^+ e^- \to Q \bar Q$
collisions in leading order where $Q$ and ${\bar Q}$ fly in opposite
directions (back-to-back).

\item In the following we assume some universal fragmentation function
\begin{equation}
D(z) = D_{Q \to H_{Q}}(z) \; ,
\end{equation}
e.g. Peterson fragmentation function often used for
$c \to D$ or $c \to \Lambda_c^{+}$ fragmentation.

\item Then, as for $e^+ e^-$ collisions, two possible prescriptions are
used:\\
   (a) $\vec{p}_{H_{Q}/pair} = z \vec{p}_{Q/pair}$ \; ,  \\
   (b) $E_{H_{Q}/pair} = z E_{Q/pair}$  \;\; and \;\; 
                     $\vec{p}_{H_{Q}/pair} || \vec{p}_{Q/pair}$ \; .
   
\item Next it is checked whether $E_{H_Q} > \frac{\sqrt{\hat{s}}}{2}$.
If not, such a case (event) is ignored.
Obviously for some $Q \bar Q$ kinematical situations the transition
$Q \to H_Q$ is not possible due to to energy-momentum conservation.
In some configurations there is not enough energy in the partonic system
to produce a heavy object.
In such a situation independent parton fragmentation makes no sense.
The fragmentation to other hadron (hadrons), coalescence, recombination
or electroweak processes can make place.

\item In the last stage four-momenta of $H_{Q}$ in the $Q \bar Q$ pair 
rest frame are transformed to the overall $p p$ or $p \bar p$ 
center-of-mass (CM) system

\item Finally distributions in rapidity, Feynman-$x_F$ and transverse
momentum of $H_Q$ in the overall $pp$ CM system are calculated.

\end{itemize}

The procedure above can, in principle, be used also for
$g \to H_Q$ or $q / {\bar q} \to H_Q$ fragmentation which is possible
in some specific cases.

To illustrate the main effects of the fragmentation we will
intentionally use collinear approach.
In the collinear approach in leading order the differential cross
section can be calculated as
\begin{equation}
\frac{d \sigma}{d y_1 d y_2 d^2 p_t} = \frac{1}{16 \pi {\hat s}^2} 
\overline{| {\cal M}_{g g \to c \bar c} |^2} x g(x_1, \mu^2) x g(x_2, \mu^2)
\; .
\end{equation}
Similar formula applies to the $g g \to g g$ (sub)process.
The relevant matrix elements are well known and can be found e.g.
in \cite{book}.
For massless outgoing partons ($gg \to gg$ process) the associated 
matrix element squared is divergent. We multiply the divergent matrix
element squared by:
\begin{equation}
F(p_t) = \frac{p_t^4}{\left( p_{t,0}^2 + p_t^2 \right)^2} \; ,
\label{Phytia_formfactor}
\end{equation}
as is routinly done e.g. in Pythia \cite{Pythia}.
The parton distributions from \cite{GJR08} will be used to
illustrate our results.

\section{Some examples}

Here we consider three examples for illustration of the proposed
in the present paper hadronization procedure:
\begin{itemize}

\item $p p \to D^0$ \; ,

\item $p p \to \eta_c$ \; ,

\item $p p \to \Lambda_c^{\pm}$ \; .

\end{itemize}

\subsection{$D$ meson production}

\begin{figure}
\includegraphics[width=8cm]{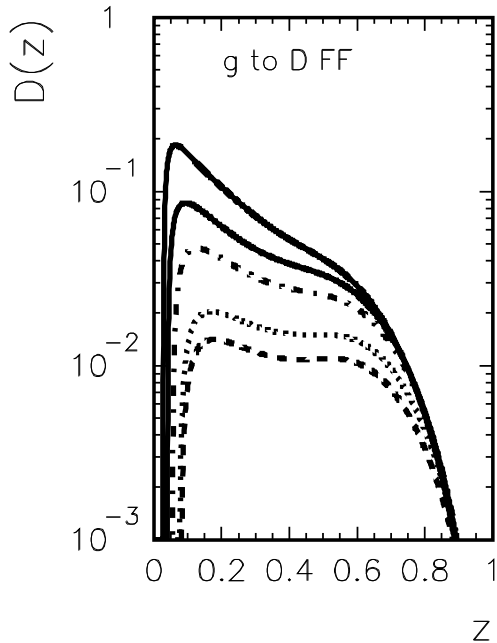}
\caption{
The $g \to D$ KKKS08 fragmentation function \cite{KKKS08} for 
different scales: 3, 4, 5, 10, 20 GeV$^2$ (from bottom to top).}
\label{fig:D_g_D}
\end{figure}

The $D$ meson production is the optimal playground for studying the
details of the fragmentation to heavy meson. Here we consider both 
$c \to D^0$ and $g \to D^0$. 
For the $c \to D^0$ the Peterson fragmentation function
is usually applied. As described in the section presenting the proposed
fragmentation procedure two methods are used for $c \to D$ fragmentation:
momentum scaling and energy scaling in the rest frame of the $c \bar c$
pair. The $g \to D$ fragmentation was discussed e.g. in \cite{KKKS08}
in the context of $e^+ e^-$ collisions.
This fragmentation function strongly depends on the evolution scale
for fragmentation (see Fig.\ref{fig:D_g_D}).

It is natural in our approach to use
$\mu_{FF}^2 = {\hat s}/4$ in analogy to the $e^+ e^-$ scattering.
The corresponding fragmentation function vanishes at the threshold.
Our choice of the evolution scale seems much more physical than
$\mu_{FF}^2 = p_t^2$ often used in the literature.
Below we shall show results of the proposed procedure for
both $c \to D$ and $g \to D$ fragmentations. 

\begin{figure}
\includegraphics[width=8cm]{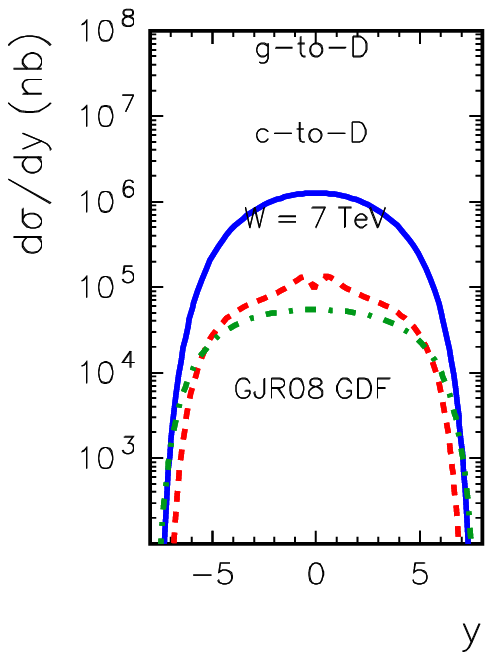}
\includegraphics[width=8cm]{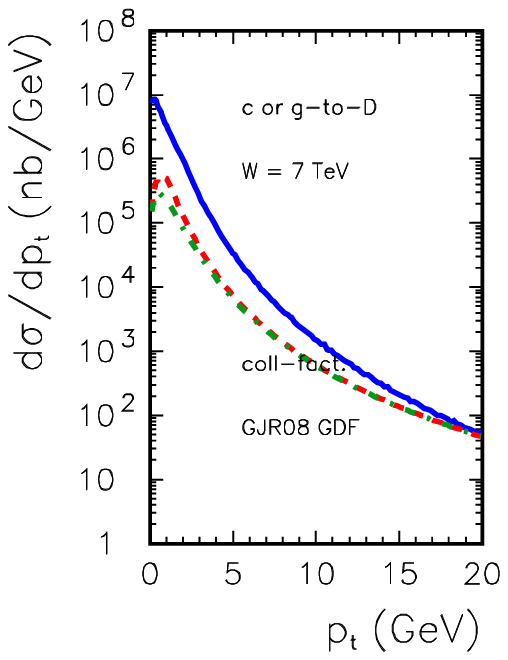}
\caption{Rapidity (left panel) and transverse momentum (right panel)
distributions for $D^0$ meson for $(c \to D^0) + (\bar c \to {\bar D}^0)$ 
and $g \to D^0 + {\bar D}^0$ fragmentation for $\sqrt{s}$ = 7 TeV. 
The fragmentation probability is included here.}
\label{fig:D_meson}
\end{figure}

In Fig.\ref{fig:D_meson} we present distributions in rapidity
of the $D^0$ meson (left panel) and corresponding distribution
in meson transverse momentum (right panel). 
The rapidity distribution for $c \to D$ fragmentation 
in the momentum scaling (red dashed line) has some
enhancement/fluctuation at $y \sim$ 0 which is a consequence of 
imposing a simple condition on energy conservation in 
the partonic subsystem. It is not the case for energy conservation
(green dash-dotted line).
Smooth distributions are obtained both for $c \to D^0$ 
(red dashed line - momentum scaling, green dash-dotted line energy scaling) 
and $g \to D^0$ (blue solid line).
We obtain a huge contribution from $g \to D^0$ but 
no singularity at $p_{t,D} \to$ 0 as in the standard approach.
The huge contribution of $g \to D^0$ is a direct consequence
of the KKKS08 parametrization. The success of the description
of experimental data for $D^0 + {\bar D}^0$ production \cite{MS2013,MS2018}
in terms of $k_t$-factorization approach combined with $c \to D^0$ 
fragmentation suggests that the normalization of 
the $g \to D^0$ fragmentation may be, in our opinion, overestimated.
The inclusion of such a $g \to D^0$ fragmentation into double $D^0 D^0$
production leads to unusually large $\sigma_{eff}$ \cite{MSSS2016},
inconsistent with other results. 
A combined fit of fragmentation functions to $e^+ e^-$ and $p p$ data 
could be useful in this context but this is not the subject of the
present exploratory study.

Now we wish to look at the Feynman-$x_F$ distributions of $D$ mesons. 
Such distributions are crucial for production of high-energy
neutrinos (see discussion in \cite{GMPS2017}). In our approach
(solid line) there is much bigger shift towards lower $x_F$ than in 
the traditional approach where $y_H = y_Q$ (dashed line) is assumed. 
We show distribution for momentum scaling (dashed line) and energy scaling
(dash-dotted line).\\
The distribution for $c$-quarks or $\bar c$-antiquarks is much broader
than that for $D$ mesons. The narrowing of the $D$ meson $x_F$-distributions
is very important for high-energy neutrino production \cite{GMPS2017}.

\begin{figure}
\includegraphics[width=8cm]{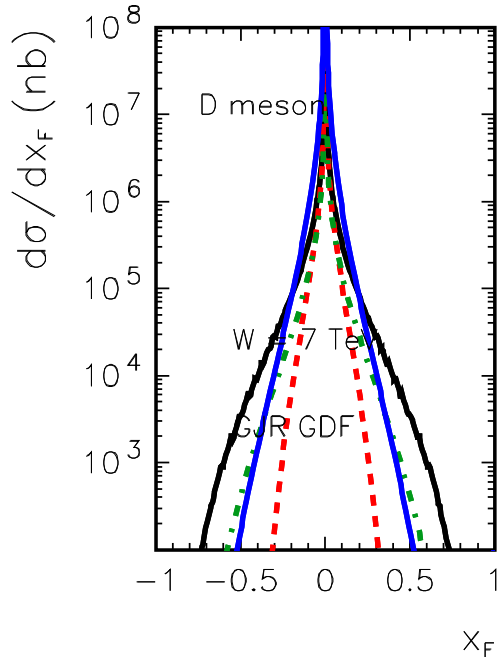}
\caption{Feynman $x_F$ distribution of charm quarks (black solid) and 
$D^0$ meson for $\sqrt{s}$ = 7 TeV. 
The red dashed line is for $(c \to D^0) + (\bar c \to {\bar D}^0)$ 
momentum scaling,
the green dash-dotted line is for $(c \to D^0) + (\bar c \to {\bar D}^0)$ 
energy scaling and
the blue solid line is for $g \to D^0 + {\bar D}^0$. 
The $c \to D^0$ fragmentation probability is included here, 
also for $c$ or ${\bar c}$ distributions.}
\label{fig:dsig_dxf_D}
\end{figure}
\subsection{$\eta_c$ production}

\begin{figure}
\includegraphics[width=8cm]{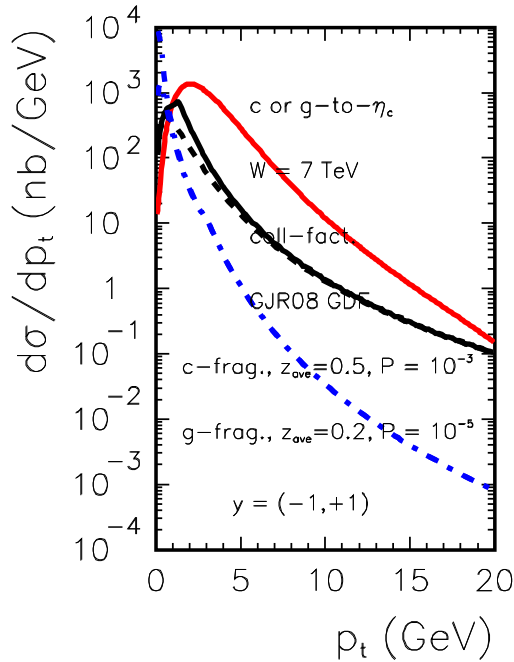}
\caption{Transverse momentum distribution of $\eta_c$ mesons
in p p collisions for $\sqrt{s}$ = 7 TeV and y = (-1,1)
for $g g \to \eta_c$ (upper solid), $c / {\bar c} \to \eta_c$
(red dahed line: momentum scaling, green dash-dotted: energy scaling)
$g \to \eta_c$ fragmentation and $g \to \eta_c$ (blue dash-dotted line).
The calculation for fragmentation mechanism was performed within
collinear factorization method using the new fragmentation method.
The GJR08 GDF was used for illustration to calculate charm and gluon 
distributions.
Here $D_{g \to \eta_c} = \delta(z-0.2)$ and
$P_{g \to \eta_c}$ = 10$^{-5}$ were used for illustration.
}
\label{fig:dsig_dpt_etac_mechanisms}
\end{figure}

The production of $\eta_c$ meson may be even a better example for 
illustration of our fragmentation procedure.
In Fig.\ref{fig:dsig_dpt_etac_mechanisms} we show transverse momentum
distribution of $\eta_c$ meson from different mechanisms 
with the focus on fragmentation contributions.
Here we show example for midrapidities $y_{\eta_c} \in$ (-1,+1)
relevant for the ALICE experiment.
For the charm-to-$\eta_c$ fragmentation the 
transition probability was roughly assumed to reproduce result
of matrix element calculation from \cite{Baranov2020}
in the midrapidity region.
Here for the gluon-to-$\eta_c$ fragmentation an approximate values of 
$P_{g \to \eta_c}$ probability and $z_{ave}$ are taken 
from Ref.\cite{Nejad2015}.

\begin{figure}
\includegraphics[width=8cm]{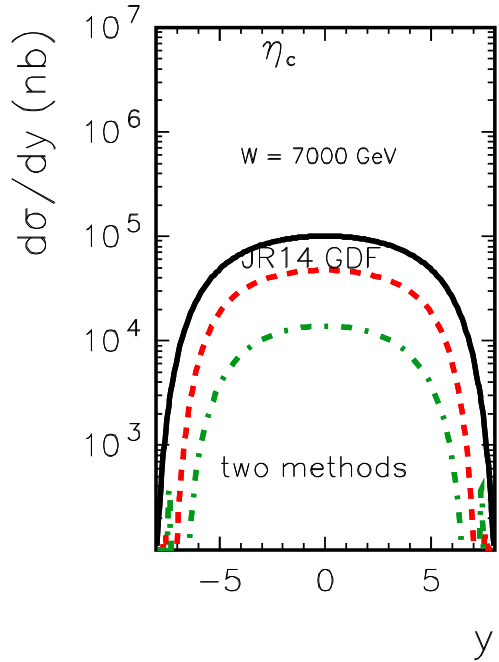}
\caption{Rapidity distribution of $\eta_c$ mesons
in p p collisions for $\sqrt{s}$ = 7 TeV for $c \to \eta_c$ fragmentation.
This calculation was performed within collinear factorization
method using the new fragmentation method.
The solid line is for momentum FF method and the dashed line is
for energy FF method. The GJR08 GDF was used to calculate $c/{\bar c}$ 
distributions.
Here $D_{c\to\eta_c} = \delta(z-0.5)$ and $P_{c \to \eta_c}$ = 10$^{-3}$
were used for illustration.
}
\label{fig:dsig_dy_etac_methods}
\end{figure}

In the standard approach distribution in rapidity of charm quarks
and mesons is the same. Within the newly proposed fragmentation scheme
it is not so. In Fig.\ref{fig:dsig_dy_etac_methods} we present
rapidity distributions of $c$ quark (solid line) and $\eta_c$ meson
(red dashed and green dash-dotted line). No $P_{c \to \eta_c}$ is 
included here, to illustrate the effect of energy conservation. 
The energy avalable in the partonic subsystem not always allow 
production of massive $\eta_c$ quarkonium, 
which leads to a sizable reduction of the cross section
compared to the partonic cross section. 
The reduction for the energy scaling (dashed line) is larger than 
the reduction for the momentum scaling (dash-dotted line).
Within the considered fragmentation scheme the rapidity distribution 
of heavy hadron ($\eta_c$) is narrower than the distribution 
of heavy quark/antiquark.

\begin{figure}
\includegraphics[width=8cm]{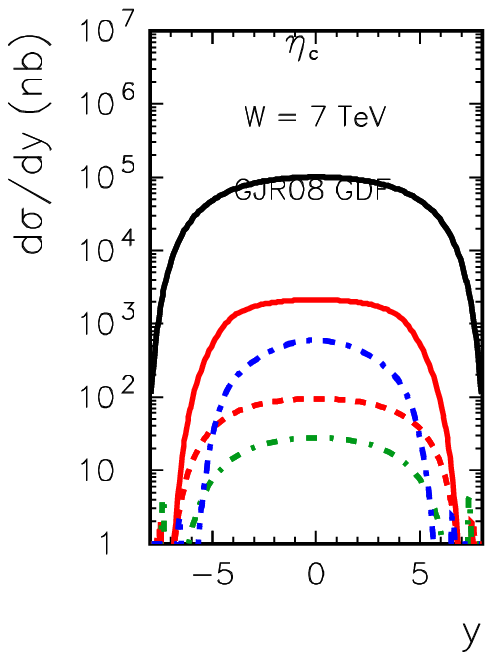}
\caption{Rapidity distribution of $\eta_c$ mesons
in p p collisions for $\sqrt{s}$ = 7 TeV for $c \to \eta_c$
and $g \to \eta_c$ fragmentation.
This calculation was performed within collinear factorization
method using new fragmentation method.
The solid line is for momentum FF method and the dashed line is
for energy FF method. The GJR08 GDF was used here for illustration.
Here $D_{c \to \eta_c} = \delta(z-0.5)$ and 
     $D_{g \to \eta_c} = \delta(z-0.2)$ 
were used for illustration and
$P_{c \to \eta_c}$ = 10$^{-3}$ and $P_{g \to \eta_c}$ = 10$^{-5}$ 
were used based on a more involved calculation \cite{Nejad2015}.
}
\label{fig:dsig_dy_etac_mechanisms}
\end{figure}

In Fig.\ref{fig:dsig_dy_etac_mechanisms} we show rapidity
distribution of $\eta_c$ from different production mechanisms.
The upper solid line is the distribution of charm quarks/antiquarks.
The lower solid line is the distribution of $\eta_c$ quarkonium
from the color-singlet gluon-gluon fusion discussed recently
in \cite{BPSS2020_etac}.
The red dashed and green dash-dotted lines correspond to 
$c \to \eta_c$ + $\bar c \to \eta_c$ for momentum and energy scaling,
respectively.
The blue dash-dotted line is for $g \to \eta_c$ fragmentation for
$\Lambda$ = 1 GeV in the regularization form factor (\ref{Phytia_formfactor}).
This distribution is much narrower than that for $c \to \eta_c$
fragmentation. We observe a huge shift $y_g \to y_{\eta_c}$ down.

\begin{figure}
\includegraphics[width=7cm]{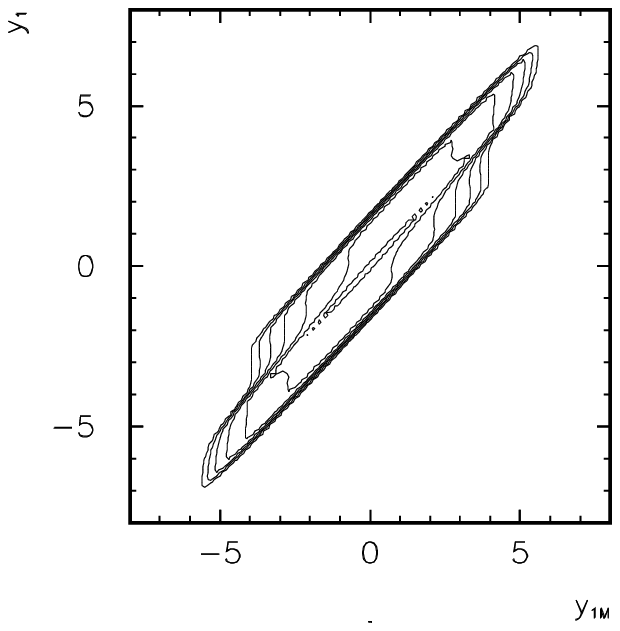}
\includegraphics[width=7cm]{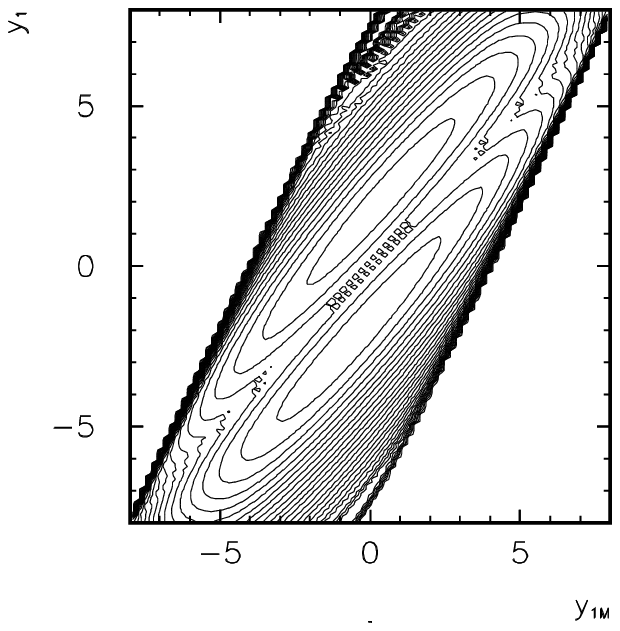}
\caption{Left panel: $(y_c, y_{\eta_c})$, 
        right panel: $(y_g, y_{\eta_c})$
for $\sqrt{s}$ = 7 TeV. 
}
\label{fig:rapidity_shift}
\end{figure}

The shift, specific for our fragmentation procedure, 
is demonstrated explicitly in Fig.\ref{fig:rapidity_shift}
for both $c \to \eta_c$ (left panel) and $g \to \eta_c$ (right panel).
In both cases we observe a shift down to smaller rapidity values. 
The shift is much bigger for gluon fragmentation than for charm
quark/antiquark fragmentation.
There is much less correlation between $y_g$ and $y_{\eta_c}$
than for $y_c$ and $y_{\eta_c}$.
It is not easy to observe the shift experimentally for $\eta_c$
production as both $c/\bar c \to \eta_c$ and the $g \to \eta_c$ 
contributions are very small compared to the color-singlet 
fusion $g^* g^* \to \eta_c$ contribution. The latter mechanism was found 
to be dominant mechanism of $\eta_c$ production 
\cite{Lansberg:2017ozx,Baranov:2019joi,BPSS2020_etac}.

\subsection{$\Lambda_c$ production}

As another example we wish to discuss production of $\Lambda_c^{\pm}$ 
in proton-proton collisions. The relevant cross sections were measured
both by the LHCb \cite{Aaij:2013mga} and ALICE 
\cite{Acharya:2017kfy} collaborations at the LHC. 
An enhanced production was observed
by the ALICE collaboration. This could not be understood in terms
of $c \to \Lambda_c^+$ parton fragmentation \cite{MS2018}.
In $e^+ e^-$ collisions at the $Z^0$ resonance the $b \to \Lambda_c^+$
fragmentation is an important ingredient \cite{OPAL}.
We wish to estimate here the $b \to \Lambda_c^+$ fragmentation in the
newly proposed fragmentation scheme.

\begin{figure}
\includegraphics[width=8cm]{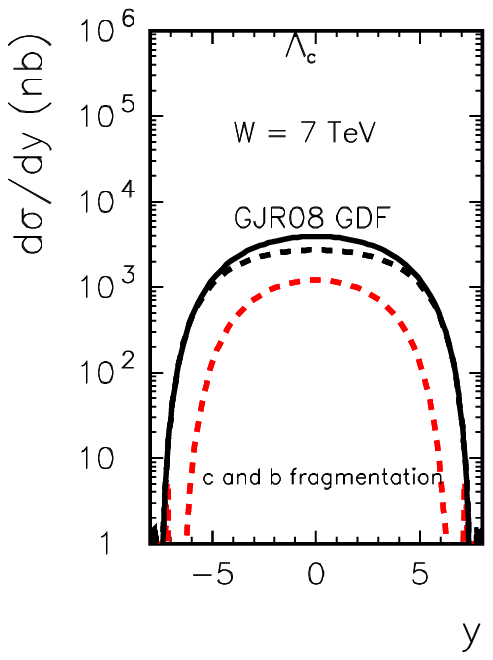}
\includegraphics[width=8cm]{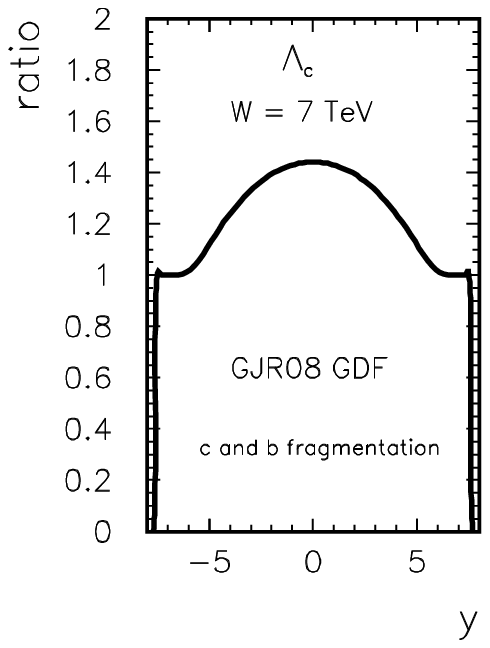}
\caption{Left panel: $d \sigma / dy$ for $\Lambda_c$ production, 
right panel: $R_{\Lambda_c}(y)$ for $\sqrt{s}$ = 7 TeV. 
}
\label{fig:Lambdac_y}
\end{figure}

In Fig.\ref{fig:Lambdac_y} we show our results for both
$c \to \Lambda_c^+$ and $b \to \Lambda_c^+$ fragmentation.
In the left panel we show rapidity distributions for $\sqrt{s}$ = 7 TeV.
The $c \to \Lambda_c^+$ contribution is shown as black dashed line,
while the $b \to \Lambda_c^+$ contribution as the red solid line. The sum
of them is shown by the black solid line. One can observe an enhancement
due to the $b \to \Lambda_c^+$ contribution at midrapidities.
The ratio:
\begin{equation}
R_{\Lambda_c}(y) = \frac{\frac{d \sigma^{c+b}}{dy}}
                         {\frac{d \sigma^{c}}{dy}}           
\label{R_y_Lambdac}
\end{equation} 
is shown in the right panel.
In this calculations we have taken $P_{c \to \Lambda_c}$ = 0.06
and $P_{b \to \Lambda_c}$ = 0.14 from \cite{KK2005} and the average
values of $z$ variable equal to 0.75 and 0.3, respectively.
The value of $P_{b \to \Lambda_c}$ seems rather upper limit in our
opinion.
Summary of branching fractions of $c$ and $b$ quarks into
charmed hadrons at LEP can be also found in \cite{Gladilin2016}.
We observe a strong dependence of the ratio on rapidity.
The relative contribution of $b \to \Lambda_c^+$ is much bigger for
midrapidities (ALICE) than for forward regions (LHCb).
In our opinion this contribution cannot, however, totally explain 
the enhancement observed by the ALICE collaboration.

\begin{figure}
\includegraphics[width=8cm]{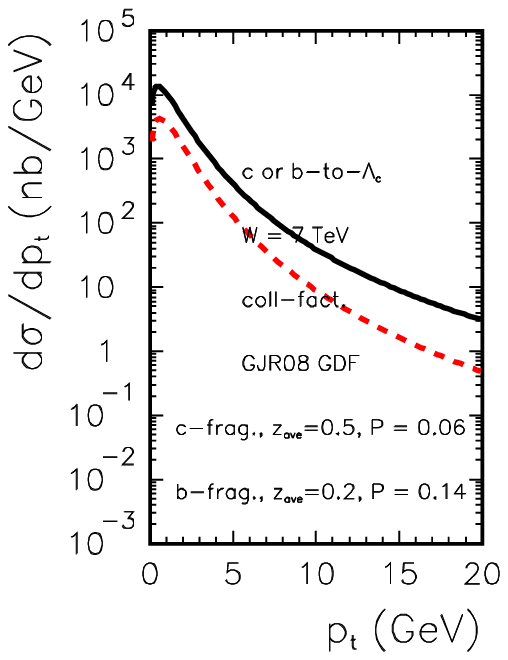}
\includegraphics[width=8cm]{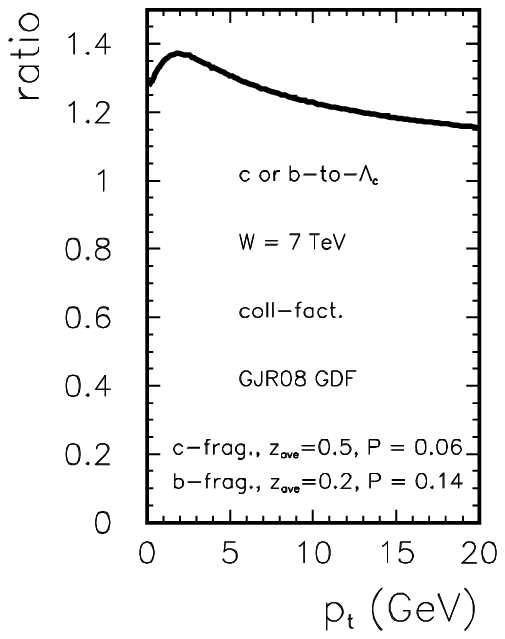}
\caption{Left panel: $d \sigma / dp_t$ for $\Lambda_c^{\pm}$ production, 
right panel: $R_{\Lambda_c}(p_t)$ for $\sqrt{s}$ = 7 TeV. 
}
\label{fig:Lambdac_pt}
\end{figure}

In Fig.\ref{fig:Lambdac_pt} we show distributions in transverse momentum
of $\Lambda_c$ and the corresponding enhancement factor.
The shape of both distributions is quite similar.

The $b \to D$ contributions are usually subtracted experimentally.
However, the statistics for $\Lambda_c^{\pm}$ was rather small and 
the experimental subtraction was not possible 
\cite{Acharya:2017kfy}.
Sometimes the nonprompt contributions is subtracted using the code
FONLL \cite{FONLL}. Of course the subtraction is model-dependent.
The proposed here model of fragmentation leads to somewhat different
rapidity dependence than in the standard approach.

\section{Conclusions}

In the present paper we have presented a new simple parton fragmentation
procedure to be applied in proton-proton collisions, more consistent with
$e^+ e^-$ collisions than the one commonly used in the literature. 
In this approach the parton-parton collision
is the basis for the hadronization. The transformation to parton-parton
center-of-mass is performed and the energy available in the partonic
subsystem is checked.
The proposed procedure applies to light parton to heavy hadron
fragmentation which is very relevant for production of mesons or baryons
containing at least one heavy quark ($c$ or $b$).

In the present paper the calculations have been performed for
leading-order collinear approach based on 2 $\to$ 2 partonic
subprocesses.
This could be generalized to $2 \to n$ processes in a future.

Some examples have been presented for illustration.
The procedure leads to a violation of the commonly used rule that
rapidity of the hadron is equal to rapidity of the parton initiating
the fragmentation. It has been shown that on average rapidity of the hadron
is smaller than rapidity of the parent parton. This is especially spectacular
for gluon-to-hadron fragmentation which leads to enhanced 
production of heavy hadrons at midrapidities. 

As an illustrative example we have considered $c \to \eta_c$
and $g \to \eta_c$ fragmentations.
We have also shown how rapidities of $\eta_c$ are correlated with rapidities
of partons ($c$ or $g$ in this case).
This leads to somewhat enhanced production of $\Lambda_c$ with respect
to $D$ mesons at midrapidities. 

We have discussed also production of $\Lambda_c^+$ via $c \to \Lambda_c^+$ 
and $b \to \Lambda_c^+$ fragmentation and found that the new procedure
would lead to enhanced production of $\Lambda_c^{\pm}$ at midrapidity.
This kinematical effect may explain
partly an enhancement observed by the ALICE collaboration 
\cite{Acharya:2017kfy}.

The proposed procedure may have also consequences for high-energy
neutrino production in the Earth's atmosphere as measured e.g.
by the Ice-Cube collaboration at the South Pole. The dominant mechanism is 
then semileptonic decays of forward produced $D$ mesons.
The distribution of forward produced $D$ mesons is then of crucial
importance. This will be discussed elsewhere.

So far we have discussed only leading-order $g g \to c \bar c$ processes.
In our opinion the procedure can be generalized to higher orders
including $g g \to c \bar c g$ or even $2 \to 4$ subprocess such as
$g g \to c \bar c g g$ or $g g \to c \bar c q \bar q$.
In those cases one should go to the $c \bar c g$ or $c \bar c g g$
center of mass systems. In such systems hadrons can be emitted in 
the direction of $c$ or $\bar c$ and the energy necessary for a
production of heavy hadron must be checked.
Final transformation to the overall center-of-mass system must be done
next to prepare final distributions in rapidity and transverse momentum.
This should be (will be) done in future.

\section*{Acknowledgements}

I would like to thank Rafa{\l} Maciu{\l}a and Sergey Baranov for a discussion. 
This study was partially supported by the Polish National
Science Center grant UMO-2018/31/B/ST2/03537 and by the Center for
Innovation and Transfer of Natural Sciences and Engineering Knowledge 
in Rzesz\'ow.

\end{document}